\RequirePackage{lineno}

\documentclass[aps,prc,showpacs,preprintnumbers,amsmath,amssymb,superscriptaddress]{revtex4}

\usepackage{graphicx} % Use pdf, png, jpg, or eps§ with pdflatex; use eps in DVI mode
\usepackage{amssymb}
\usepackage{dcolumn}% Align table columns on decimal point
\usepackage{bm}% bold math
\usepackage{bbm}% bold math
\usepackage{dsfont}% bold math
\usepackage{color}%to color the text background
\newcommand{ \be }{\begin{linenomath*}\begin{eqnarray}}
\newcommand{ \ee }{\end{eqnarray}\end{linenomath*}}
\def \mean#1 {{\left\langle #1 \right\rangle}}

\definecolor{dgreen}{cmyk}{1.,0.,1.,0.2}        % dark green
\definecolor{orange}{cmyk}{0.,0.353,1.,0.}    % orange

\begin{document}
\title{Differential Correlation Measurements with the Identity Method}
\author{Claude A. Pruneau}
\affiliation{Department of Physics and Astronomy, Wayne State University, Michigan, USA}
\author{Alice Ohlson}
\affiliation{Physikalisches Institut, Universitaet Heidelberg, Germany}
\date{\today}

\begin{abstract}
We present an extension of the identity method initially introduced for particle yield fluctuation studies  towards measurements of differential correlations. The extension is developed and illustrated in the context of measurements of the normalized two-particle cumulant $R_2$ but is  adaptable to 
any correlation measurements, including differential flow measurements.   The identity method is also extended to account for an arbitrary number of particle identification devices and signals.
\end{abstract}
\maketitle

\section{Introduction}

Studies of integral and differential correlation functions of elementary particles produced in high-energy nucleus-nucleus collisions provide invaluable information on the particle production dynamics and the collision system evolution,  and might also enable the determination of fundamental properties of the quark matter produced in these collisions~\cite{PhysRevLett.81.4816,PhysRevLett.95.182301,PhysRevC.72.064903}. Such measurements have been carried out for different 
collision systems, several beam energies, and a host of particle combinations~\cite{Voloshin:1999yf,Pruneau:2002yf,Khachatryan:2014jra,Chatrchyan:2011eka}. Semi-exclusive correlation functions measured for specific particle 
species (e.g., pions, kaons, protons, etc)  are of particular interest as they probe the influence 
of specific particle production processes determined by quantum number conservation laws. For 
instance, extensive measurements of general balance functions should provide 
detailed probes of the formation, evolution, and hadronization of the quark matter 
produced in relativistic heavy-ion collisions~\cite{PhysRevLett.85.2689,PhysRevC.65.044902,Pratt:2011bc}. The difficulty arises, however, in that such measurements of correlation functions require large datasets and severe particle rejection may be experimentally  incurred to achieve  high species purity and low contamination. Indeed, traditional  methods if selecting the species of interest and rejecting contaminating species are based on cuts on particle identification signals and typically often end up throwing away a significant fraction of the measured particles, or severely limiting the kinematic range of the measurement, or both. However, the identity method~\cite{Gazdzicki:1997gm} provides a technique to essentially  recover the full statistics and extend the kinematic range of measurements while providing reliable disambiguation  of particle species. The technique was first proposed for measurements of 
the first and second moments of particle multiplicities (integral correlation functions) with 
 two particle species but was successively extended to handle an arbitrary number of species, 
higher moments~\cite{Gorenstein:2011hr,Rustamov:2012bx}, and measurements of moments in the presence of transverse momentum-dependent efficiency losses~\cite{Pruneau:2017fim}. The method is extended here to measurements of differential correlation functions, more specifically measurements  of the normalized two-particle cumulants, $R_2$. However, the method  can be  extended to other types of two-particle correlators or to multi-particle correlation functions. The method is developed for an arbitrary  number of particle species  and accounts for particle losses due to finite detector efficiency. It is also extended to account for two or more particle identification signals. 

This paper is divided as follows. Section~\ref{sec:Correlation Function Definition} defines the normalized
two-particle differential cumulant $R_2^{(p,q)}$, for particle species $p$ and $q$,  and summarizes a 
technique, introduced elsewhere~\cite{Ravan:2013lwa},  to discretize and correct 
measurements  for particle losses. Section~\ref{sec:IdentityMethod} builds on the identity method described in Refs.~\cite{Gazdzicki:1997gm,Rustamov:2012bx,Gorenstein:2011hr,Pruneau:2017fim} and its extension involving 
an explicit dependence on detection efficiencies, towards measurements of multiplicity moments as a function of relative rapidity and  differences in azimuthal angle. Section~\ref{sec:IM2more} discusses an extension of the identity method for measurements involving more than one source of particle identification, e.g.,  studies involving joint measurements of energy loss and time-of-flight.
This work is summarized in Sec.~\ref{sec:Summary}. 

\section{$R_2$ Definition and Evaluation Technique}
\label{sec:Correlation Function Definition}

Measurements of normalized two-particle cumulants, $R_2^{(p,q)}(\Delta\eta,\Delta\phi)$, where $p$ and $q$ represent particle species  in specific kinematic ranges while $\Delta\eta$ and $\Delta\phi$ represent rapidity (or pseudorapidity) and azimuthal angle differences; triggered correlation functions $\frac{1}{N_{\rm trig}}\frac{d^2N_{\rm pairs}}{d\Delta\eta d\Delta\phi}$; and balance functions, $B(\Delta\eta,\Delta\phi)$;  have been carried out  in various shapes or forms  for a wide range of collision systems and beam energies~\cite{PhysRevC.82.024905,Abelev2013267,Abelev2010239,PhysRevLett.90.172301,Adam:2015gda,Adamczyk:2015yga}.  Physical properties and   several measurement techniques of $R_2$ were reported in~\cite{Ravan:2013lwa}. The correlator $R_2^{(p,q)}$ is commonly measured as a function of the relative rapidity (or pseudorapidity),  the difference of azimuthal angles of produced particles, or both. However, in this paper, following Method 2 of Ref.~\cite{Ravan:2013lwa}, one defines $R_2^{(p,q)}$ in four dimensions in  terms of  single- and two-particle densities, noted $\rho_1^{(p)}(y,\phi)$ and $\rho_2^{(p,q)}(y_1,\phi_1, y_2,\phi_2)$, respectively, according to:
\be\label{eq:R2-4D}
R_2^{(p,q)}(y_1,\phi_1, y_2,\phi_2) = \frac{\rho_2^{(p,q)}(y_1,\phi_1, y_2,\phi_2) }{\rho_1^{(p)}(y_1,\phi_1)\rho_1^{(q)}(y_2,\phi_2)} - 1,
\ee
where $y_i$ and $\phi_i$ (for $i=1,2$) are the rapidity (or pseudorapidity) and azimuthal angle of measured particles. The correlation function is readily reduced to  a function of the relative rapidity $\Delta y= y_1 - y_2$ and the azimuthal angle difference $\Delta \phi = \phi_1 - \phi_2$ by averaging across the measurement acceptance
\be\label{eq:R2-4D}
R_2^{(p,q)}(\Delta y,\Delta\phi) = \frac{1}{\Omega(\Delta y)}\int_{\Omega} R_2^{(p,q)}(y_1,\phi_1, y_2,\phi_2) \delta(\Delta y-y_1+y_2)\delta(\Delta \phi- \phi_1+\phi_2) dy_1 d\phi_1 dy_2 d\phi_2,
\ee
where $\Omega(\Delta y)$ represents the width of the acceptance in $\bar y = \frac{1}{2}(y_1+y_2)$ for a given value of $\Delta y$, and the relative angle $\Delta \phi$ is calculated modulo $2\pi$.

The above expression holds for continuous density functions. In practice, the experimental evaluation of $R_2^{(p,q)}$ is based on histograms  with finite size bins and the evaluation of the above integral is formulated as a discrete sum~\cite{Ravan:2013lwa} of single and pair  yields 
measured as a function of rapidity (or pseudorapidity), azimuthal angle, and transverse momentum ($p_{\rm \perp}$). In general, the measurement may be carried out with arbitrarily many bins
in all three dimensions for both single particles and pairs of particles, and for species $p$ and $q$. It is thus convenient to define three-dimensional histograms $H_1^{(p)}(\vec\alpha)$ and six-dimensional histograms $H_2^{(p,q)}(\vec \alpha,\vec \beta)$ for measurements of single 
and pair densities, respectively. The three-dimensional  vectors  $\vec \alpha=(\alpha_y,\alpha_{\phi},\alpha_{p_{\rm \perp}})$ and $\vec \beta=(\beta_y,\beta_{\phi},\beta_{p_{\rm \perp}})$ represent bin indices in rapidity (pseudorapidity), azimuthal angle, and transverse momentum. The number of bins along each axis, denoted by $m_y$, $m_{\phi}$, and $m_{p_{\rm \perp}}$, and range of the variables are to be chosen 
considering the physics of interest, the  available statistics, and the dependence of the detection efficiency on these variables.  Considering a selected data sample consisting of $N_{\rm ev}$ events, the analysis involves processing all events and  counting numbers of single  particles and  pairs into  single and pair histograms according to their respective momentum vectors, $\vec \alpha$ and $\vec \beta$.   In the absence of (or neglecting) particle losses, statistical estimators  of the single and pair densities are obtained  according to 
\be\label{eq:Density1}
\hat \rho_1^{(p)}(\vec \alpha) &\equiv& \frac{\left\langle N_{p}(\vec \alpha)\right\rangle}{\delta y \delta \phi \delta p_{\rm \perp}}, \\ 
\label{eq:Density2}
\hat \rho_2^{(p,q)}(\vec \alpha,\vec \beta) &\equiv& \frac{ \left\langle N_p(\vec \alpha)\left[ N_p(\vec \beta)-\delta_{p,q}\delta_{\vec \alpha,\vec \beta} \right] \right\rangle}{\delta y^2 \delta \phi^2 \delta p_{\rm \perp}^2 },   
\ee
where quantities $\left\langle O\right\rangle$ are event ensemble  averages of the (single or pair)  yields  in  momentum bins $\vec \alpha$ ($\vec \beta$) of width $\delta y$, $\delta \phi$, and  $\delta p_{\rm \perp}$ in  rapidity, azimuthal angle, and transverse momentum, respectively. The hats (e.g., $\hat \rho$) denote the fact that the above quantities are statistical estimators of the single and pair densities towards which they converge in the large statistics limit and for infinitesimal bin widths.

In order to obtain measurements of two-particle correlation functions in terms of the particle separation
in rapidity $\Delta y$ and azimuth $\Delta \phi$, one first sums over the $p_{\rm \perp}$ indices $\alpha_{p_{\rm \perp}}$ and $\beta_{p_{\rm \perp}}$ to obtain densities that are functions of rapidity and angle exclusively:
\be
\hat \rho_1^{(p)}(\vec \alpha^{(2)}) &=& \sum_{\alpha_{p_{\rm \perp}}=1}^{m_{p_{\rm \perp}}} \hat \rho_1^{(p)}(\vec \alpha); \hspace{0.5in} 
\hat \rho_2^{(p,q)}(\vec \alpha^{(2)},\vec \beta^{(2)})
=  \sum_{\alpha_{p_{\rm \perp}},\beta_{p_{\rm \perp}}=1}^{m_{p_{\rm \perp}}} \hat \rho_2^{(p,q)}(\vec \alpha,\vec \beta),
\ee
where $\vec \alpha^{(2)}=(\alpha_y,\alpha_{\phi})$ and $\vec \beta^{(2)}=(\beta_y,\beta_{\phi})$.
The four-dimensional normalized cumulant $R_2^{(p,q)}(y_1,\phi_1, y_2,\phi_2)$ is then evaluated according to 
\be\label{eq:4D-R2}
R_2^{(p,q)}(\vec \alpha^{(2)},\vec \beta^{(2)}) = \frac{\hat \rho_2^{(p,q)}(\vec \alpha^{(2)},\vec \beta^{(2)})}{\hat \rho_1^{(p)}(\vec \alpha^{(2)}) \hat \rho_1^{(q)}(\vec \beta^{(2)})} - 1.
\ee
Finally, $R_2^{(p,q)}$ is obtained in terms of  rapidity and azimuthal angle differences according to 
\be\label{eq:4D-R2-AVG}
R_2^{(p,q)}(\vec{\Delta\alpha}) = \frac{1}{\Omega(\Delta\alpha_y)} \sum_{\alpha_y,\alpha_{\phi},\beta_y,\beta_{\phi}} R_2^{(p,q)}(\vec \alpha^{(2)},\vec \beta^{(2)}) 
\delta(\Delta\alpha_y - \alpha_y + \beta_y)
\delta(\Delta\alpha_{\phi} - \alpha_{\phi} + \beta_{\phi}),
\ee
where the index $\Delta\alpha_y$ corresponds to rapidity difference bins, $\Delta y$, in the range $y_{\min} \le y < y_{\max}$
and the index $\Delta\alpha_{\phi}$ corresponds to  azimuthal difference bins, $\Delta \phi$, in the range
$0 \le \phi < 2\pi$, while $\Omega(\Delta\alpha_y)$ is a normalization constant that accounts for the width of the experimental acceptance in $\bar y = \frac{1}{2}(y_1+y_2)$ at a given $\Delta y$. The sums are taken over all rapidity and azimuthal bins and the delta functions insure that the differences of rapidity (angle) bins are properly matched to the $\Delta y$  ($\Delta \phi$)
bins represented by $\vec {\Delta\alpha}$. Note that the above integer arithmetic 
yields some bin sharing (often termed aliasing). This bin sharing can be modeled and corrected for or suppressed by oversampling. The bin sharing has modest effects as long as the cumulant changes slowly with $\Delta y$  and $\Delta \phi$.

Equations~(\ref{eq:Density1})--(\ref{eq:Density2}) express unbiased estimators of the densities $\rho_1^{(p)}$ and $\rho_2^{(p,q)}$ in the absence of particle losses and contamination from secondary particles or feed-down decays. The strength of the background associated with secondary particles may be evaluated with 
various track quality criteria, e.g., by applying a selection criterion on the distance of closest approach of charged  tracks to the collision primary vertex, while contributions from feed-down may require modeling of such decays. In  the context of the extension of the identity method to measurements of differential correlation functions presented in this work, the  focus is on the effects of particle losses. To this end, one must first describe the calculation of the moments of the multiplicities in bins 
$\vec \alpha$ and $\vec \beta$ in the presence of fluctuations associated with particle losses.

Proceeding similarly as in Ref.~\cite{Pruneau:2017fim}, one describes  fluctuations in the particle production according to a hypothetical (true) joint probability distribution $P_{\rm T}(\vec N_1, \vec N_2,\ldots, \vec N_K)$, in which
$\vec N_1, \vec N_2, \ldots, \vec N_K$ represent vectors of the (produced) multiplicity of particles of species $p=1, \ldots, K$ in momentum-space bins $\vec \alpha \equiv (\alpha_y,\alpha_{\phi},\alpha_{p_{\rm \perp}})$ where  $\alpha_y=1, \ldots, m_y$; $\alpha_{\phi}=1,\ldots, m_{\phi}$; and $\alpha_{p_{\rm \perp}}=1, \ldots, m_{p_{\rm \perp}}$. It is also convenient to define vectors $ \vec n_p$ and $\vec \varepsilon_p$
corresponding to vectors of measured multiplicities and detection efficiencies (defined later in this section). One can then write
\be\label{eq:XXXXXXXX}
\vec n_p&=& \left(n_p(1,1,1),n_p(1,1,2),\ldots, n_p(m_{y},m_{\phi},m_{p_{\rm \perp}})\right),  \\
\vec N_p&=& \left(N_p(1,1,1),N_p(1,1,2),\ldots, N_p(m_{y},m_{\phi},m_{p_{\rm \perp}})\right), \\
\vec \varepsilon_p &=& \left(\varepsilon_p(1,1,1),\varepsilon_p(1,1,2),\ldots, \varepsilon_p(m_{y},m_{\phi},m_{p_{\rm \perp}})\right).
\ee
Moments of the  multiplicities $N_p(\vec \alpha)$ are   calculated 
according to 
\be
\left\langle N_p(\vec \alpha)\right\rangle &=& \sum_{\vec N} N_p(\vec \alpha) P_{\rm T}(\vec N_1, \vec N_2,\ldotp, \vec N_K), \\ 
\left\langle N_p(\vec \alpha)\left[N_q(\vec \beta) -\delta_{p,q}\delta_{\vec \alpha,\vec \beta} \right] \right\rangle &=& \sum_{\vec N} N_p(\vec \alpha)\left[ N_p(\vec \beta) - \delta_{p,q}\delta_{\vec \alpha,\vec \beta} \right] P_{\rm T}(\vec N_1, \vec N_2,\ldotp, \vec N_K), \\ 
%\left\langle N_p(\vec \alpha)N_q(\vec \beta)\right\rangle &=& \sum_{\vec N} N_p(\vec \alpha)N_q(\vec \beta) P_{\rm T}(\vec N_1, \vec N_2,\ldotp, \vec N_K),
\ee
where the shorthand notation $\sum_{\vec N}$ is defined according to 
\be 
\sum_{\vec N} = \sum_{N_1(1,1,1)=0}^{\infty} \cdots \sum_{N_1(m_y,m_{\phi},m_{p_{\rm \perp}})=0}^{\infty}
\sum_{N_2(1,1,1)=0}^{\infty}\cdots \sum_{N_2(m_y,m_{\phi},m_{p_{\rm \perp}})=0}^{\infty}
\cdots
\sum_{N_K(1,1,1)=0}^{\infty}\cdots \sum_{N_K(m_y,m_{\phi},m_{p_{\rm \perp}})=0}^{\infty}
\ee
Experimentally, measurements of particle production are subjected to random losses of particles. Assuming the detection of the $N$ particles  amounts to $N$ independent processes, i.e., provided that the probability of detecting the $N$ particles jointly is equal to the product of the probabilities of detecting each of the particles independently,  one models the particle detection process in  bin $\vec \alpha$ according to a  binomial distribution $B(n_p(\vec \alpha)|N_p(\vec \alpha),\varepsilon_p(\vec \alpha))$ defined according to 
\be
\label{eq:BinomialDist}
B(n|N,\varepsilon) = \frac{N!}{n! (N-n)!} \varepsilon^{n} \left( 1 - \varepsilon\right)^{N-n},
\ee
where $\varepsilon_p(\vec \alpha)$ represents the detection efficiency of particle species $p$ 
in phase-space bin $\vec \alpha$, while $n_p(\vec \alpha)$ and $N_p(\vec \alpha)$ are the measured and true particle multiplicities in that bin.  In general, detection efficiencies differ for species $p=1,\ldots, K$ and  may also feature   dependences on $y$, $\phi$, and $p_{\perp}$, represented here as discretized functions  $\varepsilon_p(\vec \alpha)$. 

The joint probability of measuring  multiplicities $n_p(\vec \alpha)$ in  bin $\vec \alpha$ is represented with a joint probability distribution, $P_M(\vec n_1, \ldots, \vec n_K)$, defined similarly as the true distribution $P_T(\vec N_1, \ldots, \vec N_K)$. For binomial efficiency sampling,
$P_M(\vec n_1, \ldots, \vec n_K)$ can be expressed in terms of the true joint probability distribution $P_T(\vec N_1, \ldots, \vec N_K)$ according to 
\be
P_M(\vec n_1, \ldots, \vec n_K) &=& \sum_{\vec N_p} P_T(\vec N_1, \ldots, \vec N_K)    \prod_{\vec \alpha_1} B(n_1(\vec \alpha_1)|N_1(\vec \alpha_1),\varepsilon_1(\vec \alpha_1)) \\ \nonumber
& &\times   \prod_{\vec \alpha_2} B(n_2(\vec \alpha_2)|N_2(\vec \alpha_2),\varepsilon_2(\vec \alpha_2)) \times
 \cdots  \times   \prod_{\vec \alpha_K} B(n_K(\vec \alpha_K)|N_K(\vec \alpha_K),\varepsilon_K(\vec \alpha_K)),
\ee
where  the shorthand notation $\prod_{\vec \alpha}$ is defined as 
\be
\prod_{\vec \alpha} = \prod_{\alpha_y=1}^{m_y} \prod_{\alpha_{\phi}=1}^{m_{\phi}}\prod_{\alpha_{p_{\rm \perp}}=1}^{m_{p_{\rm \perp}}} 
\ee
The first and second order moments of $n_p$ are   calculated according to 
\be
\left\langle n_p(\vec \alpha) \right\rangle &=& \sum_{\vec n} n_p(\vec \alpha) P_M(\vec n_1, \ldots, \vec n_K) \\ 
\left\langle n_p(\vec \alpha)\left[ n_q(\vec \alpha)-\delta_{p,q}\delta_{\vec \alpha,\vec \beta}\right] \right\rangle &=& \sum_{\vec n} n_p(\vec \alpha)\left[ n_p(\vec \beta)-\delta_{p,q}\delta_{\vec \alpha,\vec \beta}\right] P_M(\vec n_1, \ldots, \vec n_K), 
%\left\langle n_p(\vec \alpha)n_q(\vec \beta) \right\rangle &=& \sum_{\vec n} n_p(\vec \alpha)n_q(\vec \beta) P_M(\vec n_1, \ldots, \vec n_K), 
\ee
where $\sum_{\vec n}$ represents sums over all particle species and all kinematic bins $\vec \alpha$ and $\vec \beta$, and 
expressions $\left\langle O\right\rangle$ once again refer to event ensemble averages of the single and pair multiplicities observed event by event in the distinct  kinematic bins $\vec \alpha$ and $\vec \beta$.

For narrow bins (but wide enough to neglect smearing and bin sharing) and binomial (efficiency) sampling, one readily verifies that the
measured single particle and  pair multiplicities  satisfy
\be
\label{eq:expMoment1}
\left\langle n_{p}(\vec \alpha)\right\rangle &=& \varepsilon_p(\vec \alpha) \left\langle N_{p}(\vec \alpha)\right\rangle, \\ 
\label{eq:expMoment2}
\left\langle n_p(\vec \alpha)\left[n_q(\vec \beta)-\delta_{p,q}\delta_{\vec \alpha,\vec \beta}\right]\right\rangle &=& \varepsilon_p(\vec \alpha) \varepsilon_q(\vec \beta) 
 \left\langle N_p(\vec \alpha)\left[N_p(\vec \beta)-\delta_{p,q}\delta_{\vec \alpha,\vec \beta}\right]\right\rangle, 
%\left\langle n_p(\vec \alpha)n_q(\vec \beta)\right\rangle &=& \varepsilon_p(\vec \alpha)\varepsilon_q(\vec \beta) \left\langle N_p(\vec \alpha)N_q(\vec \beta)\right\rangle.
\ee
Evidently, if the joint detection of particles in bins $(\vec \alpha)$ and $(\vec \beta)$ is correlated, one must replace the products
$\varepsilon_p(\vec \alpha)\varepsilon_q(\vec \beta)$ by  true pair efficiencies $\varepsilon_{pq}(\vec \alpha,\vec \beta)$. In general, however, one finds pair efficiencies  factorize to a good approximation and measurements of the $R_2^{(p,q)}$ correlation function in six dimensions are thus in principle inherently robust against single particle losses associated with detector or track reconstruction 
algorithm artifacts~\cite{monica} given, for instance, 
\be
R_2^{\rm M}(\vec \alpha,\vec \beta) &=& 
\frac{ \left\langle n_{p}(\vec \alpha) n_{q}(\vec \beta)\right\rangle}{ \left\langle n_{p}(\vec \alpha)\right\rangle\left\langle  n_{q}(\vec \beta)\right\rangle} - 1 = \frac{ \varepsilon_p(\vec \alpha)\varepsilon_q(\vec \beta)\left\langle N_{p}(\vec \alpha) N_{q}(\vec \beta)\right\rangle}
{\varepsilon_p(\vec \alpha) \left\langle N_{p}(\vec \alpha)\right\rangle \varepsilon_q(\vec \beta)\left\langle N_{q}(\vec \beta)\right\rangle} - 1, \\ 
&=& \frac{ \left\langle N_{p}(\vec \alpha) N_{q}(\vec \beta)\right\rangle}
{  \left\langle N_{p}(\vec \alpha)\right\rangle  \left\langle N_{q}(\vec \beta)\right\rangle} - 1 \equiv  R_2^{\rm T}(\vec \alpha,\vec \beta),
\ee
where $R_2^{\rm M}(\vec \alpha,\vec \beta)$ and $R_2^{\rm T}(\vec \alpha,\vec \beta)$ represent the measured and true normalized 
cumulants, respectively.
In practice, however, a measurement in six dimensions is challenging because at high transverse momentum, the number of particles observed in a given bin $\vec \alpha$
may be too small to enable a meaningful evaluation of $R_2$ with the above expression. Rather than calculating the ratio in six dimensions, it is more practical and common to first integrate the single and pair densities  in transverse momentum to obtain a measurement of $R_2$ in four dimensions, as in Eq.~(\ref{eq:4D-R2}),  with subsequent averaging over the acceptance to obtain a measurement as a function of $\Delta y$ and $\Delta \phi$, as in Eq.~(\ref{eq:4D-R2-AVG}). 

Using Eqs.~(\ref{eq:expMoment1})--(\ref{eq:expMoment2}), one writes:
\be
\label{eq:momentVsEpsilon1}
\left\langle  N_{p}(\vec \alpha^{(2)})\right\rangle &=& \sum_{\alpha_{p_{\rm \perp}}=1}^{m_{p_{\rm \perp}}} \frac{\left\langle  n_{p}(\vec \alpha)\right\rangle}{\varepsilon_p(\vec \alpha)}, \\ 
\label{eq:momentVsEpsilon2}
\left\langle  N_{p}(\vec \alpha^{(2)}) \left[ N_{p}(\vec \beta^{(2)})-\delta_{p,q}\delta_{\vec \alpha,\vec \beta}\right]\right\rangle &=& 
 \sum_{\alpha_{p_{\rm \perp}},\beta_{p_{\rm \perp}}=1}^{m_{p_{\rm \perp}}} 
\frac{\left\langle n_{p}(\alpha_y,\alpha_{\phi},\alpha_{p_{\rm \perp}})\left[n_{p}(\beta_y,\beta_{\phi},\beta_{p_{\rm \perp}})- \delta_{p,q}\delta_{\vec \alpha,\vec \beta}\right]\right\rangle}{\varepsilon_p( \alpha_y,\alpha_{\phi},\alpha_{p_{\rm \perp}})\varepsilon_p( \beta_y,\beta_{\phi},\beta_{p_{\rm \perp}})},
\ee
where  $\vec \alpha^{(2)} = (\alpha_y,\alpha_{\phi})$, $\vec \beta^{(2)} = (\beta_y,\beta_{\phi})$. Division by efficiencies nominally corrects for non-uniform particle losses across the detector acceptance.  Note that it is here assumed that pair efficiencies factorize into products of single efficiencies. This may not be appropriate if the momentum bins are very narrow thereby corresponding 
to detection configurations in which tracks may nearly or fully overlap (e.g., in a time projection chamber) or 
share many common detection units (e.g., in segmented tracking chambers). For measurements of pair correlations within such narrow bins, it is then more
appropriate to divide by a pair efficiency that accounts for pair losses due to partial or full track overlaps. Either way, the 
measured normalized cumulants, corrected for efficiencies,  become 
\be\label{eq:normalizedCumulant4D}
R_2^{\rm{M}(p,q)}(\vec \alpha^{(2)},\vec \beta^{(2)}) &=&
\frac{\left\langle  N_{p}(\vec \alpha^{(2)}) \left( N_{q}(\vec \beta^{(2)})-\delta_{p,q}\delta_{\vec \alpha,\vec \beta}\right)\right\rangle }
{\left\langle  N_{p}(\vec \alpha^{(2)})\right\rangle \left\langle  N_{q}(\vec \beta^{(2)})\right\rangle} -1, 
\ee
where $p,q=1,\ldots,K$ and $\vec \alpha^{(2)}$ and $\vec \beta^{(2)}$ represent arbitrary kinematic bins in the acceptance of the measurement.

Correlated losses and efficiency dependences on detection geometry (e.g., dependence of the efficiency on the collision vertex position), accelerator luminosity, detector occupancy, etc,  can be handled with vertex position or luminosity dependent weights~\cite{method_prc}. Such effects are neglected in the discussion that follows but are relatively straightforward to implement. It should be noted in closing this section that the dimensionality reduction achieved in Eqs.~(\ref{eq:momentVsEpsilon1})--(\ref{eq:momentVsEpsilon2}) can be trivially extended to yield correlation functions that are functions of $\Delta y$ or $\Delta \phi$ only, or even integral correlations yielding measures of multiplicity fluctuations such as those discussed in Ref.~\cite{Pruneau:2017fim}. 

\section{Differential $R_2$ measurements with the identity method}
\label{sec:IdentityMethod}

Studies of $R_2$ (and similar observables) have been conducted for a variety of collision systems and beam energies, various momentum ranges, and for a wide range of particle pair types ranging from inclusive charged particles to specific charge combinations, and even specific particle species. Measurements of $R_2^{(p,q)}$ for specific particle species $p$ and $q$, e.g., pions ($\pi^{\pm}$), kaons ($K^{\pm}$), or protons ($p$ or $\bar p$), are of particular interest as they provide more detailed information about the particle production process than semi-exclusive single particle measurements or inclusive correlation measurements. They may also be combined to obtain charge dependent correlations, balance functions, and general balance functions that may 
further our understanding of particle production dynamics in nuclear collisions. In the context of traditional measurements, charged particle species are identified with cuts on particle identification (PID) signals  from a time projection chamber (TPC), a time-of-flight system (TOF), etc.  Unfortunately, with such techniques, the necessity to properly disambiguate particle species typically implies the measured kinematic ranges must be limited to regions of good PID separation thereby  leading to potentially substantial particle losses.  Using the  identity method, however, 
one can recover most of  the statistics lost with conventional cut methods and significantly extend 
the kinematic range of an analysis. The method  was introduced in Ref.~\cite{Gazdzicki:1997gm} for two particle species, $p=1,2$, extended in Ref.~\cite{Rustamov:2012bx,Gorenstein:2011hr} for $K>2$ species, i.e., for $p,q=1,\ldots, K$, and the determination of higher moments, and further
 extended in Ref.~\cite{Pruneau:2017fim} to explicitly account for $p_{\rm \perp}$-dependent detection efficiencies. In this and the next section, one shows that the efficiency-dependent identity method~\cite{Pruneau:2017fim} can be further extended to differential correlation functions, such as $R_2$, provided one 
 discretizes single and pair densities according to Eqs.~(\ref{eq:Density1})--(\ref{eq:Density2}). The method presented in this 
 section relies on a single PID variable, e.g., energy loss in a time projection chamber. It is extended to measurements involving  two or more PID signals in the following section.

Within the identity method, rather than attempting to unambiguously identify
the species of measured particles event-by-event, one  relies on a 
probabilistic evaluation of the  moments $\left\langle n_k\right\rangle$ and $\left\langle n_k(n_k-1)\right\rangle$.
Specifically, instead of summing integer counts (1 for an identified particle, 0 otherwise),  one accounts for ambiguities by summing  weights $\omega_k(m)$ for each PID hypothesis. The weights are determined particle-by-particle for each hypothesis $k=1,\ldots,K$,  according to the relative frequency of particles of type $k$ for a PID signal of amplitude  $m$ (the ``mass'' signal) defined by 
\be\label{eq:pidProb}
\omega_k(m) \equiv \frac{\rho_k(m)}{\rho(m)},
\ee
with 
\be\label{eq:pidProb2}\rho(m)\equiv\sum_{k=1}^K \rho_k(m); \hspace{1in}
\int  \rho_k(m)dm = \left\langle N_k\right\rangle,
\ee
where $\rho_k(m)$ represents the number density of the PID signal $m$ for particles of type $k$ and $\rho(m)$ is the ensemble averaged PID signal density. The weight $\omega_k(m)$ expresses the probability  a PID signal of amplitude $m$ is generated by a particle of species $k$.

The goal of this work  is  to formulate  differential correlations as  functions of particle pair separation
in rapidity and azimuthal angle using the identity method. It is important to first establish that the experimentally measured signal line shape can be meaningfully used to determine the relative probability of particle species on an event-by-event basis. As an example, one considers the energy loss signal $dE/dx$  produced
by charged particles in a TPC. The momentum space is discretized in $m_y$ rapidity bins, $m_{\phi}$ azimuthal angle bins, and $m_{p_{\rm \perp}}$ transverse momentum bins. The detector response is thus expressed in terms of the discretized momentum vectors $\vec \alpha$ and $\vec \beta$ as defined in the previous section.

Let $P(k, \vec \alpha)$ represent the probability of a particle of type $k=1,\ldots,K$ being produced in momentum bin $\vec \alpha$. Further define  
$P(d|k, \vec \alpha)\equiv\varepsilon_k(\vec \alpha)$  as the conditional probability of the predicate  $d$ stating that a particle of type $k$ and momentum $\vec \alpha$ is detected in the TPC, and $P(m|d,k, \vec \alpha)$, the conditional probability density that this particle, being detected, produces a PID signal of amplitude $m$. The joint probability of 
having a particle of type $k$ being detected in the TPC and producing a signal of amplitude $m$ is thus
\be\label{eq:joint1}
P(m,d,k, \vec \alpha) =  P(m|d,k, \vec \alpha) \varepsilon_k(\vec \alpha) P(k, \vec \alpha),
\ee
where we substituted $\varepsilon_k(\vec \alpha)$ for $P(d|k, \vec \alpha)$.
We use  $P(m,d,k, \vec \alpha)$ to   calculate  the probability that a signal of amplitude $m$ corresponds to a particle of type $k$:
\be\label{eq:joint2}
P(m,d,k, \vec \alpha) = P(k|m,d,\vec \alpha) P(m|d,\vec \alpha)P(d,\vec \alpha),
\ee
where $P(k|m,d,\vec \alpha)$ represents the conditional probability that a track detected in the TPC 
with a PID signal of amplitude $m$ and  momentum-space coordinate bin $\vec \alpha$   corresponds to a particle of type $k$; $P(m|d,\vec \alpha)$ represents the conditional probability density that a PID signal of amplitude $m$ be produced 
when a particle within the momentum-space bin $\vec \alpha$ is detected in the TPC and $P(d,\vec \alpha)$ represents the joint probability a particle of momentum $\vec \alpha$ be observed in the TPC. Using Eqs.~(\ref{eq:joint1})--(\ref{eq:joint2}), one writes (Bayes' theorem) 
\be
P(k| m,d,\vec \alpha) = \frac{  P(m|d,k, \vec \alpha)\varepsilon_k(\vec \alpha) P(k, \vec \alpha)}{P(m|d,\vec \alpha)P(d,\vec \alpha)}.
\ee
 The quantity $\varepsilon_k(\vec \alpha)$   represents the detection efficiency of 
 particles of type $k$ at  momentum $\vec \alpha$ and  can be determined by Monte Carlo 
 simulations of the detector performance or by embedding techniques.   
 $P(m|d,k, \vec \alpha)$ represents the line shape of the PID signal $m$ associated with 
 a detected particle of type $k$ (it corresponds to $\omega_k(m)$ in Eq.~(\ref{eq:pidProb})), whereas 
 $P(k, \vec \alpha)=P(k|\vec \alpha)P(\vec \alpha)$ corresponds to the joint
 probability, determined statistically from the event ensemble average, that 
 a produced particle of momentum $\vec \alpha$ and type $k$ are detected. The quantity
 $P(m|d,\vec \alpha)$ represents the probability that a PID signal $m$  is observed when a 
 particle at momentum $\alpha$ is detected, while 
 $P(d,\vec \alpha)$ represents the  joint probability that a particle be detected in the TPC at 
 a momentum $\vec \alpha$. $P(m|d,\vec \alpha)$  is obtained by summing the probability densities 
of  PID signal $m$  associated with all species 
 \be
P(m|d,\vec \alpha) P(d,\vec \alpha) &=& \sum_{k=1}^K P(m|d,k,\vec \alpha)\varepsilon_{k}(\vec \alpha) P(k, \vec \alpha) 
 \ee 
and 
\be
P(d,\vec \alpha) = \sum_{k=1}^K P(d|k, \vec \alpha)P(k, \vec \alpha)  = \sum_{k=1}^K \varepsilon_{k}(\vec \alpha) P(k, \vec \alpha).
\ee
The overall line shape $P(m|d,\vec \alpha)=\rho(m)/\left\langle N\right\rangle$ is   given by
\be
P(m|d,\vec \alpha) = \frac{\sum_{k=1}^K P(m|d,k, \vec \alpha)\varepsilon_{k}(\vec \alpha) P(k, \vec \alpha)}
{\sum_{k=1}^K \varepsilon_{k}(\vec \alpha) P(k, \vec \alpha)}
\ee
The conditional probability $P(k| m,d,\vec \alpha)$ can then be expressed
\be\label{eq:probabilityOfType}
P(k| m,d,\vec \alpha) = \frac{  P(m|d,k,\vec \alpha)\varepsilon_{k}(\vec \alpha)P(k, \vec \alpha)}
{\sum_{k'=1}^{K} P(m|d,k',\vec \alpha) \varepsilon_{k'}(\vec \alpha) P(k',\vec \alpha)}
\ee
One finally obtains  the line shape $ \rho_k(m|\vec \alpha)$ for particles of type $k$ in the momentum bin $\vec \alpha$:
\be
 \rho_k(m|\vec \alpha) =   P(m|d,k, \vec \alpha) \varepsilon_{k}(\vec \alpha) P(k, \vec \alpha) \left\langle N(\vec \alpha) \right\rangle,
\ee
where $\left\langle N(\vec \alpha) \right\rangle = \sum_k \left\langle N_k(\vec \alpha) \right\rangle$
One thus finds that, indeed,  Eq.~(\ref{eq:probabilityOfType}) is equivalent to Eq.~(\ref{eq:pidProb}), and 
$\omega_k(m)$  corresponds to the probability of species $k$ given a PID signal of amplitude $m$ at a specific momentum $\vec \alpha$, which one thus denotes
\be
\omega_k(m|\vec \alpha) = \frac{  P(m|d,k, \vec \alpha) \varepsilon_{k}(\vec \alpha) P(k, \vec \alpha)}
{\sum_{k'=1}^{K}  P(m|d,k',\vec \alpha)\varepsilon_{k'}(\vec \alpha) P(k',\vec \alpha)}
\ee
The weights $\omega_k(m|\vec \alpha)$ provide the correct
probability of a particle being of species $k$ given $m$  only if they are evaluated
as a function of the momentum vector $\vec \alpha$. Indeed, the relative probability of 
species $k=1,\ldots,K$ may be a function of rapidity, azimuthal angle, and transverse momentum. In practice, it may be unnecessary to use the same level of granularity 
for the determination of the weights $\omega_k(m|\vec \alpha)$ and
the study of the particle densities. This is particularly important in the context of experiments where efficiencies depend on the collision centrality (or event multiplicity), the collision vertex position, or any other additional variables.

Following the original  identity method, one  defines an event-by-event quantity $W_p$, hereafter called an event-wise identity variable, for species $p=1,\ldots, K$, as the sum of the  weights $\omega_p(m|\vec \alpha)$ over all $M$ particles in an event which satisfy the kinematic and quality criteria used in the analysis:
\be\label{eq:IdentityVariable}
W_p(\vec \alpha) \equiv \sum_{i=1}^{M} \omega_p(m_i|\vec \alpha).
\ee
The identity method   involves  calculating the moments of $W_p(\vec \alpha)$ and
we shall verify that they are linear combinations of the moments of $N_p(\vec \alpha)$.
For measurements of $R_2^{(p,q)}$, one only  needs to consider the two lowest moments
\be\label{eq:wp}%%%%%%%%%
\left\langle W_p(\vec \alpha)\right\rangle &=& \frac{1}{N_{\rm events}} \sum_{i=1}^{N_{\rm events}}W_p^{(i)}(\vec \alpha), \\ 
\label{eq:wp2}
\left\langle W_p(\vec \alpha) W_q(\vec \beta)\right\rangle&=& \frac{1}{N_{\rm events}} \sum_{i=1}^{N_{\rm events}}W_p^{(i)}(\vec \alpha)W_q^{(i)}(\vec \beta).
\ee
in which $W_p^{(i)}(\vec \alpha)$ and $W_q^{(i)}(\vec \beta)$  are event-wise identity variables for species $p$ and $q$  in events $i=1,\ldots, N_{\rm events}$, measured in kinematic bins   $\vec \alpha$ and $\vec \beta$, respectively.

Theoretically, calculations of the expectation values of the moments $\left\langle W_p(\vec \alpha)\right\rangle$
and $\left\langle W_p(\vec \alpha)W_q(\vec \beta)\right\rangle$ with particle losses proceed  similarly as in Ref.~\cite{Pruneau:2017fim}, but  one must properly average over all species, all bins $\vec \alpha$, and all particles in those bins. The resulting mathematical expressions are rather large and cumbersome; it is thus convenient to develop some additional shorthand notations.
Given that one must account for binomial sampling in each bin $\vec \alpha$, for each species $p$, let us introduce
\be\label{eq:bp}
\mathbb{B}(\vec n_p,\vec N_p, \vec \varepsilon_p) =
 \prod_{\alpha_y=1}^{m_{y}}
 \prod_{\alpha_{\phi}=1}^{m_{\phi}}
 \prod_{\alpha_{p_{\rm \perp}}=1}^{m_{p_{\rm \perp}}} B(n_p(\vec \alpha)|N_p(\vec \alpha), \varepsilon_p(\vec \alpha)),
\ee
where $\vec n_p$,  $\vec N_p$, $\vec \varepsilon_p$ represent vectors of values in all bins $\vec \alpha=(\alpha_y,\alpha_{\phi},\alpha_{p_{\rm \perp}})$ introduced in Eq.~(\ref{eq:XXXXXXXX}).

One must also average over all possible values of PID signals, for all species, in all bins 
$\vec \alpha$. To that end, one defines functionals 
\be
\mathbb{P}_p(n_p(\vec \alpha)) = \prod_{i=1}^{n_p(\vec \alpha)} 
\int P(m_i|d,p,\vec \alpha) dm_i,
\ee
where $n_p(\vec \alpha)$ is the number of particles of species $p$ detected in bin $\vec \alpha$; $m_i$ is the amplitude of the PID signal of  the $i$-th particle of type $p$ in that bin; and $P(m_i|d,p,\vec \alpha)$ is the probability density of such signals. In order to average over all bins $\vec \alpha$, one introduces the functionals
\be
\mathbb{S}_p(\vec n_p) =  
 \prod_{\alpha_y=1}^{m_{y}}
 \prod_{\alpha_{\phi}=1}^{m_{\phi}}
 \prod_{\alpha_{p_{\rm \perp}}=1}^{m_{p_{\rm \perp}}} \mathbb{P}_p(n_p(\vec \alpha))
\ee
The integrals within the functionals  $\mathbb{P}_p(n_p(\vec \alpha))$ and 
$\mathbb{S}_p(\vec n_p)$ are to be evaluated when multiplied to the right by $W_p$. The expectation 
value of $W_p(\vec \alpha)$ may then be written 
\be\label{eq:Wxxxx}
\left\langle W_p(\vec \alpha)\right\rangle = \sum_{\vec N} \sum_{\vec n} P_T(\vec N) 
\prod_{k=1}^K \mathbb{B}(\vec n_k,\vec N_k, \vec \varepsilon_k) \mathbb{S}_p(\vec n_p)
\sum_{k'=1}^K \sum_{i_{k'}=1}^{n_{k'}(\vec \alpha)} \omega_p(m_{i_{k'}}^{(k')}|\vec \alpha)) 
\ee
This expression involves products of several integrals whose evaluation seems daunting. However, note that most of the integrals are of  the form $\int P(m)dm=1$ and thus do not contribute to $\left\langle W_p(\vec \alpha)\right\rangle$. Only integrals of the form $\int \omega_p(m) P(m|d,q,\vec \alpha) dm$
 yield non-unitary values and must thus be accounted for. Similarly as in Ref.~\cite{Pruneau:2017fim}, it is convenient to introduce response coefficients
\be\label{eq:rpq}
r_{pq}(\vec \alpha)&=& \int \omega_p(m|\vec \alpha) P(m|d,q,\vec \alpha)dm, \ee
Equation~(\ref{eq:Wxxxx}) may then be written
\be
\left\langle W_p(\vec \alpha)\right\rangle &=& \sum_{\vec N} \sum_{\vec n} P_T(\vec N)
  \prod_{k=1}^K \mathbb{B}_k(\vec n_p,\vec N_p, \vec \varepsilon_p)
  \sum_{k'=1}^K  r_{pk'}(\vec \alpha) n_{k'}(\vec \alpha).
  \ee
  Sequential evaluation of the sums $\sum_{\vec n}$ and $\sum_{\vec N}$ yields
\be
\left\langle W_p(\vec \alpha)\right\rangle &=& \sum_{\vec N}  P_T(\vec N)
  \sum_{k=1}^K  r_{pk}(\vec \alpha) N_{k}(\vec \alpha) \varepsilon_k(\vec \alpha)\\  \nonumber
   &=& \sum_{k=1}^K r_{pk}(\vec \alpha) \left\langle N_{k}(\vec \alpha)\right\rangle \varepsilon_k(\vec \alpha).
\ee
As in Ref.~\cite{Pruneau:2017fim}, it is convenient to absorb the efficiencies into the moments and write
\be\label{eq:wwwww}
\left\langle W_p(\vec \alpha)\right\rangle &=&\sum_{k=1}^K r_{pk}(\vec \alpha) \left\langle n_{k}(\vec \alpha)\right\rangle,
\ee
where, by definition, $\left\langle n_{k}(\vec \alpha)\right\rangle =  \left\langle N_{k}(\vec \alpha)\right\rangle \varepsilon_k(\vec \alpha)$. For a given bin $\vec \alpha$, the above equation expresses the 
averages $\left\langle W_p(\vec \alpha)\right\rangle$ as a linear combination of the average multiplicities
$\left\langle n_{k}(\vec \alpha)\right\rangle$ determined by the coefficients $r_{pk}(\vec \alpha)$.
One then introduces vectors 
\be
\mathbb{\vec W}(\vec \alpha) &\equiv&  \left(\left\langle W_1(\vec \alpha)\right\rangle,\left\langle W_2(\vec \alpha)\right\rangle, \ldots, \left\langle W_K(\vec \alpha)\right\rangle \right),  \\ 
\mathbb{\vec N}(\vec \alpha) &\equiv& \left(\left\langle n_1(\vec \alpha)\right\rangle,\left\langle n_2(\vec \alpha)\right\rangle, \ldots, \left\langle N_K(\vec \alpha)\right\rangle\right),   
\ee
and the response matrices 
\be
\mathbb{R}(\vec \alpha ) = \left( {\begin{array}{*{20}{c}}
  {{r_{11}}(\vec \alpha )}& \cdots &{{r_{1K}}(\vec \alpha )} \\ 
   \vdots & \ddots & \vdots  \\ 
  {{r_{K1}}(\vec \alpha )}& \cdots &{{r_{KK}}(\vec \alpha )} 
\end{array}} \right).
\ee
The $K$ equations in~(\ref{eq:wwwww}) may then be written
\be\label{eq:wwwwwInv}
\mathbb{\vec W}(\vec \alpha) = \mathbb{R}(\vec \alpha) \mathbb{\vec N}(\vec \alpha).
\ee
The average multiplicities $\mathbb{\vec N}(\vec \alpha)$ are thus obtained by inversion of $\mathbb{R}(\vec \alpha )$:
\be
\mathbb{\vec N}(\vec \alpha) = \left(\mathbb{R}(\vec \alpha )\right)^{-1} \mathbb{\vec W}(\vec \alpha),
\ee
and average multiplicities corrected for efficiency losses, $\left\langle N_p(\vec \alpha) \right\rangle$, are then calculated, for each species $p=1,\ldots,K$, according to
\be
\left\langle N_p(\vec \alpha) \right\rangle = \frac{\left\langle n_p(\vec \alpha) \right\rangle }{\varepsilon_p(\vec \alpha)}.
\ee
Note that there are $m_y \times m_{\phi} \times m_{p_{\rm \perp}}$ independent matrix inversions to carry out, one for each momentum bin $\vec \alpha$. If momentum smearing was an important effect, one would have to invoke  smearing response functions and all these matrix inversions would be coupled.

Evaluation of the second order moments proceeds similarly. However, 
one must consider separately the four cases corresponding to Eq.~(\ref{eq:wp2}): 
$\left\langle W_p(\vec \alpha)^2\right\rangle$, $\left\langle W_p(\vec \alpha)W_p(\vec \beta)\right\rangle$, $\left\langle W_p(\vec \alpha)W_q(\vec \alpha)\right\rangle$, 
and $\left\langle W_p(\vec \alpha)W_q(\vec \beta)\right\rangle$,
with $p\ne q$ and $\vec \alpha\ne \vec \beta$. Toward that end, it is convenient to define
\be
\label{eq:rpqj}
r_{pqk}(\vec \alpha) &=& \int \omega_p(m|\vec \alpha)\omega_q(m|\vec \alpha) P(m|d,k,\vec \alpha)dm, 
\ee
The second and cross moments, $\left\langle W_p(\vec \alpha)W_q(\vec \beta) \right\rangle$, are calculated according to  
\be
\left\langle W_p(\vec \alpha)W_q(\vec \beta)\right\rangle &=& \sum_{\vec N} \sum_{\vec n} P_T(\vec N) 
\prod_{k=1}^K \mathbb{B}(\vec n_k,\vec N_k, \vec \varepsilon_k) \mathbb{S}_k(\vec n_k)
\left[\sum_{k'=1}^K \sum_{i_{k'}=1}^{n_{k'}} \omega_p(m_{i_{k'}}^{(k')}|\vec \alpha))\right]^2 , \\ 
%%%
&=& \sum_{k=1}^K  r_{pqk}(\vec \alpha) \left\langle N_{k}(\vec \alpha)\right\rangle  \varepsilon_{k}(\vec \alpha) \delta_{\vec\alpha,\vec\beta}\\ \nonumber
& & + \sum_{k,k'=1}^K  r_{pk}(\vec \alpha) r_{qk'}(\vec \beta) \left\langle N_{k}(\vec \alpha) \left[N_{k'}(\vec \beta) -  \delta_{k,k'} \delta_{\vec\alpha,\vec\beta}  \right\rangle \right] \varepsilon_{k}(\vec \alpha)\varepsilon_{k'}(\vec \alpha).
 \ee
The efficiencies can be reabsorbed within the average multiplicities and number of pairs. The above expression simplifies to
\be\label{eq:wppaa}
\left\langle W_p(\vec \alpha)W_q(\vec \beta)\right\rangle &=& 
\sum_{k=1}^K  r_{pqk}(\vec \alpha) \left\langle n_{k}(\vec \alpha)\right\rangle \delta_{\vec\alpha,\vec\beta}   \\ \nonumber
%& &  + \sum_{k=1}^K \left[ r_{pk}(\vec \alpha)\right]^2  \left\langle n_{k}(\vec \alpha) \left[n_{k}(\vec \alpha) - 1 \right] \right\rangle \\ \nonumber 
& & + \sum_{k, k'=1}^K r_{pk}(\vec \alpha)  \left\langle n_{k}(\vec \alpha)\left[ n_{k'}(\vec \beta)- \delta_{k,k'} \delta_{\vec\alpha,\vec\beta} \right] \right\rangle r_{qk'}(\vec \beta), 
\ee
where $p,q=1,\ldots,K$, while $\vec\alpha$ and $\vec\beta$ represent arbitrary kinematic bins.
It is  useful to define the matrices 
 \be\label{eq:NMat2}
\mathbb{N}(\vec \alpha,\vec \beta ) = 
\left[ {\begin{array}{*{20}{c}}
  {{N_{11}}(\vec \alpha ,\vec \beta )}&{{N_{12}}(\vec \alpha ,\vec \beta )}& \cdots &{{N_{1K}}(\vec \alpha ,\vec \beta )} \\ 
  {{N_{21}}(\vec \alpha ,\vec \beta )}&{{N_{22}}(\vec \alpha ,\vec \beta )}& \cdots &{{N_{2K}}(\vec \alpha ,\vec \beta )} \\ 
   \vdots & \vdots & \ddots & \vdots  \\ 
  {{N_{K1}}(\vec \alpha ,\vec \beta )}&{{N_{K2}}(\vec \alpha ,\vec \beta )}& \cdots &{{N_{KK}}(\vec \alpha ,\vec \beta )} 
\end{array}} \right].
\ee
with elements 
\be
N_{pq}(\vec \alpha, \vec \beta) &=& \left\langle n_p(\vec \alpha )\left[ n_q(\vec \beta ) - \delta_{p,q}\delta_{\vec\alpha,\vec\beta} \right] \right\rangle
\ee
and   
\be\label{eq:vMat2}
\mathbb{V}(\vec \alpha,\vec \beta ) = 
\left[ {\begin{array}{*{20}{c}}
  {{V_{11}}(\vec \alpha ,\vec \beta )}&{{V_{12}}(\vec \alpha ,\vec \beta )}& \cdots &{{V_{1K}}(\vec \alpha ,\vec \beta )} \\ 
  {{V_{21}}(\vec \alpha ,\vec \beta )}&{{V_{22}}(\vec \alpha ,\vec \beta )}& \cdots &{{V_{2K}}(\vec \alpha ,\vec \beta )} \\ 
   \vdots & \vdots & \ddots & \vdots  \\ 
  {{V_{K1}}(\vec \alpha ,\vec \beta )}&{{V_{K2}}(\vec \alpha ,\vec \beta )}& \cdots &{{V_{KK}}(\vec \alpha ,\vec \beta )} 
\end{array}} \right].
\ee
with elements 
\be
V_{pq}(\vec \alpha, \vec \beta) &=& \left\langle W_p(\vec \alpha)W_q(\vec \beta)\right\rangle  
- \sum_{k=1}^K   r_{pqk}(\vec \alpha) \left\langle n_{k}(\vec \alpha)\right\rangle \delta_{\vec\alpha,\vec\beta}, 
\ee
Equation~(\ref{eq:wppaa}) can then be written in matrix form 
\be
\mathbb{V}(\vec \alpha,\vec \beta) &=& \mathbb{R}(\vec \alpha)\mathbb{N}(\vec \alpha,\vec \beta)\mathbb{R}(\vec \beta)^T.
\ee
Multiplying on the left and on the right by the  inverses of matrices $\mathbb{R}(\vec \alpha)$ and $\mathbb{R}(\vec \beta)^T$, one gets
\be
\label{eq:NvsUVU}
\mathbb{N}(\vec \alpha,\vec \beta) &=&\mathbb{R}(\vec \alpha)^{-1}\mathbb{V}(\vec \alpha,\vec \beta)\left(\mathbb{R}(\vec \beta)^T\right)^{-1}.
\ee
This expression corresponds to a set of $(m_y \times m_{\phi} \times m_{p_{\rm \perp}})^2$ independent equations, one for each pair of bins $\vec\alpha$ and $\vec\beta$.  The matrices $\mathbb{N}(\vec \alpha,\vec \beta)$ can thus be calculated independently for each pair $\vec\alpha,\vec\beta$. The elements of these matrices then yield the  second and cross moments of the  multiplicities, $\left\langle n_p(\vec \alpha)\left[ n_q(\vec \beta) -\delta_{\vec\alpha,\vec\beta}\delta_{p,q} \right]\right\rangle$. The above formulation in terms of matrices $\mathbb{V}$
involves a significant and convenient simplification of the inversion problem as it was first 
presented in Ref.~\cite{Pruneau:2017fim}. 

Estimates of the true second moments, corrected for efficiency losses, 
are finally obtained according to
\be
\left\langle N_p(\vec \alpha)\left[ N_q(\vec \beta) -\delta_{\vec\alpha,\vec\beta}\delta_{p,q} \right]\right\rangle &=& \frac{\left\langle n_p(\vec \alpha)\left[ n_q(\vec \beta) -\delta_{\vec\alpha,\vec\beta}\delta_{p,q} \right]\right\rangle }{\varepsilon_p(\vec \alpha)\varepsilon_q(\vec \beta)}. 
\ee
Again in this case, if two-particle efficiencies do not properly factorize into products of single particle efficiencies, estimates of two-particle efficiencies can be used in the above in lieu of the products of singles.

The matrix inversion technique outlined above  provides second moments of particle multiplicities corrected for efficiency across the fiducial acceptance of the experiment. It is worth noting, however, that while in the above formulation, the matrices are small (determined by the number of species), there can be  many of them to invert. For instance, for an analysis involving a rapidity acceptance $-1 \le y\le 1$ in 20 bins, full azimuthal acceptance in 72 bins, and 20 bins in $p_{\rm \perp}$, one would need 28,800 matrices. This is evidently not an issue from a computational standpoint with modern computers, but it does have two practical implications. First, the available statistics will be distributed across many bins and it is conceivable that the number of entries in a given bin and the corresponding statistical uncertainty may yield numerically unstable results. Additionally, since the coefficients $r_{pq}$ are based on global fits of the line shapes in each kinematic bin $\alpha$ (although, to reiterate, the granularity required for such fits can likely be coarse), it might be necessary to manually inspect all the fits and make sure they are not subject to idiosyncrasies of the analysis or the detector performance. Differential analyses with the identity method thus clearly have high computing and storage costs.

Finally, note that once the moments $\left\langle N_p(\vec \alpha)\left[ N_q(\vec \beta) -\delta_{\vec\alpha,\vec\beta}\delta_{p,q} \right]\right\rangle$   are obtained for some nominal range of transverse momentum, i.e.,  given $m_{p_{\rm \perp}}$ bins  for $\alpha_{p_{\rm \perp}}$ 
and $\beta_{p_{\rm \perp}}$, one can readily 
obtain sums $\left\langle N_p(\vec \alpha^{(2)})\left[ N_q(\vec \alpha^{(2)}) -\delta_{\vec\alpha,\vec\beta}\delta_{p,q} \right]\right\rangle$ including all $m_{p_{\rm \perp}}$ bins or only a restricted  range of $p_{\rm \perp}$ using Eq.~(\ref{eq:momentVsEpsilon2}). It is thus possible to compare 
results obtained with the identity method described here with those obtained with traditional cut methods (applicable only over a limited range of transverse momentum) by selecting appropriate $p_{\rm \perp}$ sum ranges for each species of interest.

\section{Identity Method with two or more identity signals}
\label{sec:IM2more}

Large collider experiments commonly feature partially redundant and complementary techniques of particle identification. For instance, the  STAR and ALICE experiments both include particle identification devices based on specific energy loss ($dE/dx$) and time of flight (TOF) measurements. The ALICE detector additionally features transition radiation detectors geared towards the identification of electrons. Bayesian identification techniques based on cuts have already been developed that exploit the joint information from several PID detectors on a track by track basis. While such techniques maximize the use of information from the multiple components of a detector, such as the ALICE detector~\cite{Kowalski:1996gra,Aamodt:2008zz,Beole:2012asa,Alessandro:2006yt,Abelev:2014ffa,Carminati:2004fp,ALICE2}, they nonetheless suffer statistical losses associated with the use of PID selection criteria. This section describes an  extension  of the identity method applied to detectors featuring  
several PID signal types available for each track.  

As a preamble to the discussion, note that PID detector components suffer efficiency losses and tracking algorithms may fail to associate a given PID detector signal to a track. In particular, there are kinematic regions in which usable $dE/dx$ and TOF signals may not be obtainable. One may thus  end up reconstructing tracks that  feature no useable $dE/dx$ signal but a reliable TOF signal,  no TOF signal but a reliable $dE/dx$ signal, or no useful PID signal at all. Since the point of the identity method is to utilize all of the available information, one needs to devise a technique to statistically include all tracks featuring  PID signals, even though the information may be  incomplete. One must thus first consider  the combination of probabilistic statements about the PID of particles.

 %%%%%%%%%
 %%%%%%%%%
In this context, one   once again uses the many probability functions (e.g., $P(p, \vec \alpha)$, $P(d|p, \vec \alpha)$) that were introduced in Sec.~\ref{sec:IdentityMethod}. However, one must also  introduce a few additional definitions and probability functions. Assume there are $N_D$ detector components potentially producing PID signals that may be associated to a track. Let $D_j$, for $j=1,\ldots,N_D$, represent the predicate ``the track is detected (or matched to a signal) in device $j$'', where one arbitrarily assigns $j=1$, for instance, to a TPC, $j=2$ to a TOF detector, and so on. Additionally, let $E_j$, for $j=1,\ldots,N_D$, represent the predicate 
 ``the PID info of device $j$ is usable.''  Finally, let $m_j$ represent the PID signals produced by devices $j=1,\ldots,N_D$. For a given track, these can be conveniently expressed as $\vec m = (m_1,m_2, \ldots, m_{N_D})$. Assuming detector topologies similar to those of  STAR and ALICE, consideration of the PID information provided by a detector component $j\ge 2$ is only meaningful if a track is first detected in device $j=1$ (e.g., a TPC track). Indeed, in the context of these experiments, the detection of a hit in the TOF detector is not useful unless it can be matched to a track from the TPC. One consequently requires that  $D_1$ be true. However, $E_1$ is not necessarily required so long as one of the other devices produces a usable PID signal, i.e., if there exists one $E_j$. 
 
 Let us first consider the predicate logic for a detection system involving two components. One shall see how it can be generalized to more than two components later in this paragraph. In the following, one indicates a true predicate by its name: $D_2$ means that a given particle is detected or matched in device $j=2$ while a barred predicate, $\bar D_2$, indicates the track is not detected or matched in device $j=2$. Using commas to denote  logical conjunctions, for a two component detection systems, only the predicate combinations  $(D_1, E_1, \bar D_2)$, $(D_1, E_1, D_2, E_2)$, $(D_1, E_1, D_2, \bar E_2)$, and $(D_1, {\bar E_1}, D_2, E_2)$ provide conditions with useful PID information.
 For instance $(D_1, E_1, \bar D_2)$ means 
 a track was detected ($D_1$) in device $j=1$, produced a usable PID signal ($E_1$) in that device, but was not detected in device $j=2$ ($\bar D_2$). Clearly, usable PID information from detector 2 ($E_2$) can only be present if there is a signal in detector 2 ($D_2$).  The alternative, $\bar{E_2}$, encompasses the case in which there is a signal in detector 2 ($D_2$) but no useable PID information as well as the cases where there is no signal in detector 2 ($\bar{D_2}$).  Therefore the information about $D_2$ and $\bar{D_2}$ is absorbed into $E_2$ and $\bar{E_2}$.   If additional PID devices are available, one needs to consider all permutations deemed appropriate. For instance, with the addition of a third device, one might have
$(D_1, E_1, E_2, E_3)$, $(D_1, \bar E_1, E_2, E_3)$, $(D_1, E_1, \bar E_2, E_3)$, 
$(D_1, E_1, E_2, \bar E_3)$, $(D_1, \bar E_1, \bar E_2, E_3)$, $(D_1, \bar E_1, E_2, \bar E_3)$,
and $(D_1, E_1, \bar  E_2, \bar E_3)$. For the sake of simplicity in the remainder of this work,  the discussion is limited to two PID devices only but extensions to $N_D>2$ are relatively straightforward.
 
The momentum and species of the particles must also be accounted for. As in  Sec.~\ref{sec:IdentityMethod}, let $k,p,q$, with $k,p,q=1,\ldots, K$,  denote  species indices (assuming $K$ distinct possibilities) and let $\vec \alpha$ and $\vec \beta$ represent momentum bin index vectors. The probability of detecting a track produced by a particle 
of species $p$ in   momentum bin $\vec \alpha$ (i.e., the efficiency)  is denoted $d_1\equiv P(D_1|p,\alpha)$. 

The probabilities that a particle produces a meaningful PID signal in both detectors ($E_1$,$E_2$), in detector 1 but not in detector 2 ($E_1$,$\bar{E_2}$), in detector 2 but not in detector 1 ($\bar{E_1}$,$E_2$), or neither detector ($\bar{E_1}$,$\bar{E_2}$) are given by  
$\varepsilon_{12}\equiv e_1e_2d_1$, 
$\varepsilon_{1}\equiv e_1(1-e_2) d_1$, 
$\varepsilon_{2}\equiv e_2(1-e_1) d_1$, and $\varepsilon_{0}\equiv (1-e_1)(1-e_2)d_1 + 1 -d_1$, respectively. Here, $e_1$ denotes the probability of having a usable PID signal in detector 1, and $e_2$ denotes  the product of the probabilities of detecting, matching, and having a useful signal in detector 2. Given an event with $N(p,\alpha)$ particles of type $p$ within the momentum bin $\vec \alpha$, the number of tracks detected with conditions  $(E_1,E_2)$,   $(E_1,\bar E_2)$,   $(\bar E_1, E_2)$, are hereafter denoted $n_{12}(p,\vec \alpha)$, $n_{1}(p, \vec \alpha)$, $n_{2}(p,\vec \alpha)$, and the number of undetected tracks (i.e., tracks not detected or those detected without a usable PID signal) is $n_{0}$. These numbers shall evidently fluctuate event by event. The probability of a given combination of the numbers is given by a multinomial probability distribution  $M(n_{12}(p,\vec \alpha),n_{1}(p,\vec \alpha),n_{2}(p,\vec \alpha)|N,\varepsilon_{12}(p,\vec \alpha),\varepsilon_{1}(p,\vec \alpha),\varepsilon_{2}(p,\vec \alpha),\varepsilon_{0}(p,\vec \alpha))$ defined according to 
\be
M(n_{12},n_{1},n_{2}|N,\varepsilon_{12},\varepsilon_{1},\varepsilon_{2},\varepsilon_{0}) = 
\frac{N!}{n_{12}!n_{1}!n_{2}!n_{0}!} \varepsilon_{12}^{n_{12}} \varepsilon_{1}^{n_{1}} \varepsilon_{2}^{n_{2}}\varepsilon_{0}^{n_{0}},
\ee
where the labels $p$ and  $\vec \alpha$ were omitted for the sake of simplicity, and $n_0= N-n_{12} -n_{1} - n_{2}$.

One must next consider the probability density distributions of signals $m_1$ and $m_2$. Assuming the generation of PID signals $m_1$ and $m_2$ are statistically independent, let $P(m_1| E_1,D_1, p,\alpha)$ and $P(m_2| E_2,D_1, p,\alpha)$ respectively represent the
probability densities of signals $m_1$ and $m_2$, with normalization $\int P(m_i| E_i,D_1, p,\alpha)dm_i=1$,  for $i=1,2$. 
PDFs expressing the probability that a measured particle is of type $p$ given PID signals of amplitude $m_i$, $i=1,2$ are obtained with Bayes' theorem
\be
P(p|m_1,m_2, E_1,E_2,D_1,\alpha) &=& \frac{P(m_1| E_1,D_1, p,\alpha)P(m_2| E_2,D_1, p,\alpha)P(E_1,E_2,D_1, p,\alpha)}{\sum_q P(m_1,m_2, E_1,E_2,D_1,q,\alpha)}, \\ 
P(p|m_1,E_1,\bar E_2,D_1,\alpha) &=& \frac{P(m_1| E_1,D_1, p,\alpha)P(E_1,\bar E_2,D_1, p,\alpha)}{\sum_q P(m_1,E_1,\bar E_2,D_1,q,\alpha)}, \\ 
P(p|m_2,\bar E_1,E_2,D_1,\alpha) &=& \frac{P(m_2| E_2,D_1, p,\alpha)P(\bar E_1,E_2,D_1, p,\alpha)}{\sum_q P(m_1,\bar E_1,E_2,D_1,q,\alpha)}, 
\ee
%%%%
It is convenient to use the shorthand notation $\vec m = (m_1, m_2)$ to define  weights according to 
\be
\omega_p^{(12)}(\vec m|\vec  \alpha) &=&P(p|m_1,m_2, E_1,E_2,D_1,\vec \alpha) ,\\\
\omega_p^{(1)}(\vec m|\vec  \alpha) &=&P(p|m_1,E_1,\bar E_2,D_1, \vec \alpha)\delta(m_2),\\ 
\omega_p^{(2)}(\vec m|\vec  \alpha) &=&P(p|m_2,\bar E_1,E_2,D_1, \vec \alpha)\delta(m_1).
\ee
where both signals $m_1$ and $m_2$ are included in all three cases for notational convenience in the following. The delta function factors $\delta(m_1)$ and $\delta(m_2)$ are included  to signify explicitly that 
the signals $m_1$ and $m_2$ are not relevant for weights $\omega_p^{(2)}(\vec m| \alpha)$ and $\omega_p^{(1)}(\vec m| \alpha)$, respectively. The weights $\omega_p^{(12)}(\vec m|\vec  \alpha)$, $\omega_p^{(1)}(\vec m|\vec  \alpha)$, and $\omega_p^{(2)}(\vec m|\vec  \alpha)$, can thus be represented
as $\omega_p^{(T)}(\vec m|\vec  \alpha)$ with types $T=(12)$, $(1)$, and $(2)$, respectively.

The event-wise identity variable $W_p$ is defined according to 
\be\label{eq:IdentityVariableMultDevices}
W_p(\vec \alpha) &=&
 \sum_{i=1}^{n_{12}} \omega_p^{(12)}(\vec m_i| \vec \alpha) 
 + \sum_{i=1}^{n_{1}} \omega_p^{(1)}(\vec m_i|  \vec \alpha) 
 + \sum_{i=1}^{n_{2}} \omega_p^{(2)}(\vec m_i| \vec \alpha),\\ 
  &=& \sum_{T}  \sum_{i_{T}=1}^{n_{T}} \omega_p^{(T)}(\vec m_{i_{T}}| \vec \alpha),
\ee
where in the first line, the three sums account for tracks satisfying $(E_1,E_2)$,   $(E_1,\bar E_2)$, and    $(\bar E_1, E_2)$, respectively, while in the second line, they were replaced with the sum $\sum_{(T)}$ which 
represents a sum (of sums) for cases $(12)$, $(1)$, and $(2)$.  

One next proceeds to calculate the expectation value of the moments  $W_p(\vec \alpha)$, and
$W_p(\vec \alpha)W_q(\vec \beta)$.  To this end, one defines  coefficients $r_{pj}^{(T)}(\vec\alpha)$ and $r_{pqj}^{(T)}(\vec\alpha)$ with $(T)=(12), (1), (2)$, which are analogs of coefficients defined by Eqs.~(\ref{eq:rpq}) and (\ref{eq:rpqj}),  according to 
\be
\label{eq:up12}
r_{pj}^{(T)}(\vec \alpha) &=& \int \omega_p^{(T)}(\vec m|\vec \alpha) P(\vec m|T,j,\vec \alpha)dm_1 dm_2, \\ 
\label{eq:rpqj-12}
r_{pqj}^{(T)}(\vec \alpha) &=& \int \omega_p^{(T)}(\vec m|\vec \alpha)\omega_q^{(T)}(\vec m|\vec \alpha) P(\vec m|T,j,\vec \alpha)dm_1 dm_2, 
\ee
where, for convenience,  one also used the shorthand $T$ within the probabilities $P(\vec m|T,j,\vec \alpha)$ to represent 
the permutations $(E_1,E_2,D_1)$, $(E_1,\bar E_2,D_1)$,$(\bar E_1,E_2,D_1)$.
In order to carry out sums on the measured particles, one needs to insert multinomial distributions in each kinematic bin. One must also average over all possible multiplicity configurations in moment space spanned by $\vec \alpha$.  One thus defines  the notation
\be\label{eq:mp}
\mathbb{M}(\vec n_p^{(12)},\vec n_p^{(1)},\vec n_p^{(2)},\vec N_p,
 \vec \varepsilon_p^{(12)},  \vec \varepsilon_p^{(1)},  \vec \varepsilon_p^{(2)}) =
 \prod_{\alpha_y=1}^{m_{y}}
 \prod_{\alpha_{\phi}=1}^{m_{\phi}}
 \prod_{\alpha_{p_{\rm \perp}}=1}^{m_{p_{\rm \perp}}} M(n_p^{(12)}(\vec \alpha),n_p^{(1)}(\vec \alpha),n_p^{(2)}(\vec \alpha)|N_p(\vec \alpha),  \vec \varepsilon_p^{(12)},  \vec \varepsilon_p^{(1)},  \vec \varepsilon_p^{(2)}),
\ee
where $\vec n_p^{(12)}$,  $\vec n_p^{(1)}$, $\vec n_p^{(2)}$ represent vectors of values of the number of particles
detected with $(E_1,E_2)$,  $(E_1,\bar E_2)$, and $(\bar E_1,E_2)$, respectively,  in all bins $\vec \alpha=(\alpha_y,\alpha_{\phi},\alpha_{p_{\rm \perp}})$. $\mathbb{M}$ expresses the probability of measurement outcomes for a given species $p$ over the full space $\vec \alpha$.  
One must also average over all possible values of PID signals, for all species, and for all bins 
$\vec \alpha$. To that end, one define functionals 
\be
\mathbb{P}_p^{(T)}(n_p^{(T)}(\vec \alpha)) &=& \prod_{k=1}^{n_p^{(T)}(\vec \alpha)} 
\int P(\vec m_{k}|T,p,\vec \alpha) d\vec m,  
\ee
and
\be
\mathbb{S}_p(\vec n_p^{(12)},\vec n_p^{(1)},\vec n_p^{(2)}) =  
 \prod_{\alpha_y=1}^{m_{y}}
 \prod_{\alpha_{\phi}=1}^{m_{\phi}}
 \prod_{\alpha_{p_{\rm \perp}}=1}^{m_{p_{\rm \perp}}} \mathbb{P}_p^{(12)}(n_p^{(12)}(\vec \alpha))\mathbb{P}_p^{(1)}(n_p^{(1)}(\vec \alpha))\mathbb{P}_p^{(2)}(n_p^{(2)}(\vec \alpha)).
\ee
The integrals within the functionals  $\mathbb{P}_p(n_p(\vec \alpha))$ and 
$\mathbb{S}_p(\vec n_p^{(12)},\vec n_p^{(1)},\vec n_p^{(2)})$ are to be evaluated when multiplied to the right by $W_p(\vec \alpha)$. The expectation 
value of $W_p(\vec \alpha)$ may then be written 
\be\label{eq:W1aa}
\left\langle W_p(\vec \alpha)\right\rangle &=&  \sum_{\vec N} \sum_{\vec n} P_T(\vec N) 
\prod_{{j'}=1}^K \mathbb{M}_{j'}(\vec n_{j'}^{(12)},\vec n_{j'}^{(1)},\vec n_{j'}^{(2)},\vec N_{j'}, \vec \varepsilon_{j'}) \mathbb{S}_{j'}(\vec n_{j'}^{(12)},\vec n_{j'}^{(1)},\vec n_{j'}^{(2)})\\ \nonumber
& & \times 
\sum_{j=1}^K \left( \sum_{T}
\sum_{i_{j}=1}^{n_{j}^{(T)}(\vec \alpha)} \omega_p^{(T)}(m_{1,i_j},m_{2,j}|\vec \alpha)) 
\right), \\
&=& \sum_{j=1}^K \left[ \sum_{T} \varepsilon_j^{(T)}(\vec \alpha) r_{pj}^{(T)}(\vec \alpha) \right] \left\langle N_{j}(\vec \alpha)\right\rangle,
\ee
where, in the last line, only the relevant integrals are kept and included in the form of the coefficients  $r_{pj}^{(T)}(\vec \alpha)$ defined in Eq.~(\ref{eq:up12}). Note that it is not possible, in this case, to
reabsorb the efficiencies into the multiplicities as in the previous section because these 
are now associated with different response coefficients $r_{pj}^{(T)}(\vec \alpha)$. One next defines the matrices $\mathbb{R}(\vec \alpha)$
with elements $R_{pj}= \sum_{T} \varepsilon_j^{(T)}(\vec \alpha) r_{pj}^{(T)}(\vec \alpha)$. The first moments  $\mathbb{\vec N}(\vec \alpha)$ are thus given by the linear equations 
\be
\mathbb{\vec N}(\vec \alpha) = \left( \mathbb{R}(\vec \alpha) \right)^{-1} \mathbb{W}(\vec \alpha).
\ee
The evaluation of the second moments and cross-moments of $W_p(\vec \alpha)$ proceeds in a similar fashion:
\be\label{eq:W2ppaa}
\left\langle W_p(\vec \alpha)W_q(\vec \beta)\right\rangle &=& 
\sum_{k=1}^K  \left[ \sum_{T}  r_{pqk}^{(T)}(\vec \alpha) \varepsilon_k^{(T)}(\vec \alpha) \right] \left\langle N_k(\vec \alpha)\right\rangle \delta_{\vec\alpha,\vec\beta} \\  \nonumber 
 &  & + \sum_{k,k'=1}^K   \left[ \sum_{T, T'}  r_{pk}^{(T)}(\vec \alpha)r_{qk'}^{(T')}(\vec \beta) 
 \varepsilon_k^{(T)}(\vec \alpha)\varepsilon_{k'}^{(T')}(\vec \beta) \right]
 \left\langle N_k(\vec \alpha)\left[N_{k'}(\vec \beta) - \delta_{k,k'}\delta_{\vec\alpha,\vec\beta} \right] \right\rangle   .
\ee 
As in the previous section, one next defines the matrices $\mathbb{V}(\vec \alpha, \vec \beta)$ with elements $V_{pq}(\vec \alpha, \vec \beta)$ calculated according to:
\be
V_{pq}(\vec \alpha, \vec \beta) &=&  \left\langle W_p(\vec \alpha)W_q(\vec \beta)\right\rangle 
- \sum_{k=1}^K \left[ \sum_{T}  r_{pqk}^{(T)}(\vec \alpha) \varepsilon_k^{(T)}(\vec\alpha) \right]  \left\langle N_k(\vec \alpha)\right\rangle  \delta_{\vec\alpha,\vec\beta}.
\ee
Equations~(\ref{eq:W2ppaa}) are then  rewritten
\be\label{eq:VW2ppaa}
V_{pq}(\vec \alpha,\vec \beta) &=& 
 \sum_{k,k'=1}^K 
 \left[  \sum_{T}  r_{pk}^{(T)}(\vec \alpha) \varepsilon_k^{(T)}(\vec \alpha)  \right] 
  \left\langle N_k(\vec \alpha)\left[N_{k'}(\vec \beta) - \delta_{k,k'}\delta_{\vec\alpha,\vec\beta} \right] \right\rangle   
  \left[  \sum_{T'}  r_{qk'}^{(T')}(\vec \beta)   \varepsilon_{k'}^{(T')}(\vec \beta) \right] 
 \ee 
Redefining the elements of  the matrices $\mathbb{N}$ and $\mathbb{R}$ according to 
\be
N_{pq}(\vec \alpha, \vec \beta) &=& \left\langle N_p(\vec \alpha )\left[ N_q(\vec \beta ) - \delta_{p,q}\delta_{\vec\alpha,\vec\beta} \right] \right\rangle,
\ee
and
\be
R_{pk}(\vec \alpha) = \sum_{T}  r_{pk}^{(T)}(\vec \alpha) \varepsilon_k^{(T)}(\vec \alpha),
\ee
one gets matrix equations
\be
\mathbb{V}(\vec \alpha,\vec \beta) &=& \mathbb{R}(\vec \alpha,\vec \beta)
\mathbb{N}(\vec \alpha,\vec \beta) \left[  \mathbb{R}(\vec \alpha,\vec \beta) \right]^T,
\ee
which are solved by multiplying on the left and right by inverses of the matrices $\mathbb{R}(\vec \alpha)$ and $\mathbb{R}(\vec \beta)^T$ thereby yielding expressions of the form of Eqs.~(\ref{eq:NvsUVU}) that provide the  moments $\left\langle N_p(\vec \alpha )\left[N_q(\vec \beta ) - \delta_{p,q}\delta_{\vec\alpha,\vec\beta} \right]\right\rangle$.
One thus concludes that, in the context of analyses involving several PID signals, the determination of multiplicity moments proceeds essentially  as in the case of a single type of PID signal. However, it is not possible, in general, to reabsorb the efficiencies in the moments because they enter in linear combinations within the coefficients $r_{pk}$. Inversion of the matrix equations thus requires both the knowledge of the functions $r_{pk}^{(T)}$ as well as that of the efficiencies $\varepsilon_k^{(T)}(\vec \alpha)$.

The above formalism was derived assuming a particular PID scheme. However, it can be adapted  to other PID requirements with little to no change to  the equations. Additionally, one could also adapt the equations so that  different PID schemes are used in different $p_{\rm \perp}$ ranges, e.g., TPC PID at low $p_{\rm \perp}$, TOF PID at high $p_{\rm \perp}$, and Cherenkov or Transition Radiation detectors in between.  

\section{Summary}
\label{sec:Summary}

A binning technique to discretize  six-dimensional two-particle correlation functions $R_2^{pq}$ was first introduced to evaluate
 two particle correlations as functions of  rapidity, azimuthal angle, and transverse momentum, and project them onto two-dimensional correlators that are functions of the particles  rapidity and azimuthal angle differences. Such discretized functions were next shown to  be amenable to measurements with the identity method first in the context of experiments with a single PID device and finally for 
 experiments featuring two  PID devices. The method  is also applicable to multiple-particle correlations and for measurement devices featuring more than 2 PID techniques. 

%%%%% acknowledgements
\newenvironment{acknowledgement}{\relax}{\relax}
\begin{acknowledgement}
\section*{Acknowledgements}

The authors thank colleague A. Rustamov for fruitful discussions and comments.
This work was supported in part by the United States Department  of Energy, Office of Nuclear Physics (DOE NP), United States of America under Grant No. DE-FOA-0001664.  This work was also supported in part by BMBF and SFB 1225 ISOQUANT, Germany.  
\end{acknowledgement}

\bibliography{imDiff_v4}

\end{document}